\documentclass[lettersize,journal]{IEEEtran}
\usepackage{amsmath,amsfonts,amsthm}
\usepackage{algorithmic}
\usepackage{algorithm}
\usepackage{array}
\usepackage[justification=justified]{caption}
\usepackage{cite}
\usepackage{float}
\usepackage{graphicx}
\usepackage{stfloats}
\usepackage{subfigure}
\usepackage{textcomp}
\usepackage{url}
\usepackage{verbatim}
\usepackage[dvipsnames]{xcolor}

\begin{document}

\title{An Approximate Wave-Number Domain Expression \\for Near-Field XL-array Channel}

\author{Hongbo Xing, Yuxiang Zhang, Jianhua Zhang, Huixin Xu, Guangyi Liu and Qixing Wang
\thanks{Hongbo Xing, Yuxiang Zhang, Jianhua Zhang and Huixin Xu are with the State Key Lab of Networking and Switching Technology, Beijing University of Posts and Telecommunications, Beijing 100876, China (e-mail: hbxing@bupt.edu.cn; zhangyx@bupt.edu.cn; jhzhang@bupt.edu.cn;  xuhuixin@bupt.edu.cn).

Guangyi Liu and Qixing Wang are with the Future Research Laboratory, China Mobile Research Institute, Beijing 100053, China (e-mail: liuguangyi@chinamobile.com; wangqixing@chinamobile.com)
}

}


\maketitle

\begin{abstract}
As Extremely large-scale array (XL-array) technology advances and carrier frequency rises, the near-field effects in communication are intensifying. 
In near-field conditions, channels exhibit a diffusion phenomenon in the angular domain, existing research indicates that this phenomenon can be leveraged for efficient parameter estimation and beam training. However, the channel model in angular domain lacks closed-form analysis, making the time complexity of the corresponding algorithm high.
To address this issue, this paper analyzes the near-field diffusion effect in the wave-number domain, where the wave-number domain can be viewed as the continuous form of the angular domain. A closed-form approximate wave-number domain expression is proposed, based on the Principle of Stationary Phase. Subsequently, we derive a simplified expression for the case where the user distance is much larger than the array aperture, which is more concise.
Subsequently, we verify the accuracy of the proposed approximate expression through simulations and demonstrate its effectiveness using a beam training example. Results indicate that the beam training scheme, improved by the wave-number domain approximation model, can effectively estimate near-field user parameters and perform beam training using far-field DFT codebooks. Moreover, its performance surpasses that of existing DFT codebook-based beam training methods.

\end{abstract}

\begin{IEEEkeywords}
near-field, Extremely large-scale array (XL-array), wave-number domain, the principle of stationary phase.
\end{IEEEkeywords}

\section{Introduction}
\IEEEPARstart{E}{xtremely} 
Large-Scale array (XL-array) stands out as a pivotal technology within the sixth generation mobile communication systems (6G) \cite{JSAC_miao}. By deploying extremely large antenna arrays in massive array, it is hoped to achieve higher spectrum efficiency and higher energy efficiency \cite{6G_vision}. 
An accurate and tractable channel model is essential for thorough exploration and utilization of XL-array \cite{3D-MIMO}. 
However, as the array aperture increases and the carrier frequency rises \cite{zhang2023channel}, the near-field region will expand\cite{raileigh-fresnel-dis}, users will be more likely to appear in the near-field region \cite{liuyuanwei_near,linglongdai_spatial}. At this point, the spherical wave assumption should be used to replace the plane wave assumption in channel models\cite{C1_Rayleigh}.

The spherical wave assumption leads to the diffusion effect in angular domain channel, resulting in the loss of sparsity in angular domain. Consequently, angular domain methods face difficulties in directly applying to channel estimation, codebook design, and beam training. Existing researches address this issue by seeking different domains. For instance, \cite{linglongdai_spatial} investigate channel estimation and near-field codebook design in the polar domain, \cite{frft} explore channel estimation in the fractional Fourier domain, and \cite{phase_mode} examine near-field codebook design in the phase mode domain. 

\setlength{\parskip}{0pt}

However, the diffusion characteristics in the angular domain can also implicitly reflect near-field channel properties, which could benefit related near-field applications. For example, \cite{youchangsheng_DFTcodebook} develop a fast near-field beam training scheme by angular domain response.
Nevertheless, due to the nonlinear characteristics of the spherical wave phase, computing the near-field angular domain channel requires non closed-form oscillatory integral, making quantitative analysis challenging.

On this basis, it is hoped to investigate a closed-form angular domain channel expression. Inspired by the similarity between spherical wave and frequency-modulated (FM) signal, we employ the Principle of Stationary Phase (POSP) \cite{POSP-for-chirp} from the field of radar signal processing to derive the approximation of near-field angular domain channel.
To analyze the characteristics of near-field angular domain channels under general conditions, we choose to conduct the analysis in the wave-number domain. The wave-number domain represents the continuous Fourier transform (FT) of continuous spatial domain channel \cite{HMIMO-wavenumber2}, where the angular domain corresponds to the discrete Fourier transform (DFT) of discrete spatial domain channel \cite{virtual_channel}. Consequently, the analysis in the wave-number domain can overcome the grid effect \cite{virtual_channel} of DFT, making it more general.

The manuscript is organized as follows. 
In Section \ref{2}, we introduce the existing channel model for XL-arrays in the spatial, angular, and wave-number domains, establishing the correspondence between the wave-number domain and the angular domain.
In Section \ref{3}, we present the approximate wave-number domain channel model based on the POSP.
In Section \ref{4}, we verify the accuracy of the approximate expression through numerical simulations. Then we demonstrate the potential applications and effectiveness of the proposed approximate model using a Beam Training Example.
Final discussions and possible extensions of this paper are set forth in Section \ref{5}.

\section{Near-field XL-array Channel Model}
\label{2}

In this section, we consider a narrow-band XL-array system. First, the existing near-field channel models in spatial domain and angular domain are introduced, then the definition of the wave-number domain is provided. Finally, the relationship between angular domain and wave-number domain are discussed.
Without loss of generality, the base station (BS) is equipped with N-antenna uniform linear array (ULA) and the user has a single-antenna. Reducing the three-dimensional spatial channel to two dimensions. Note that it is possible to extend the present model into a three-dimensional model.\cite{zhang_ele}

\subsection{Spatial and Angular Domain Near-field Channel Model}

\begin{figure}[htbp]
\centering
\includegraphics[width=0.4\textwidth]{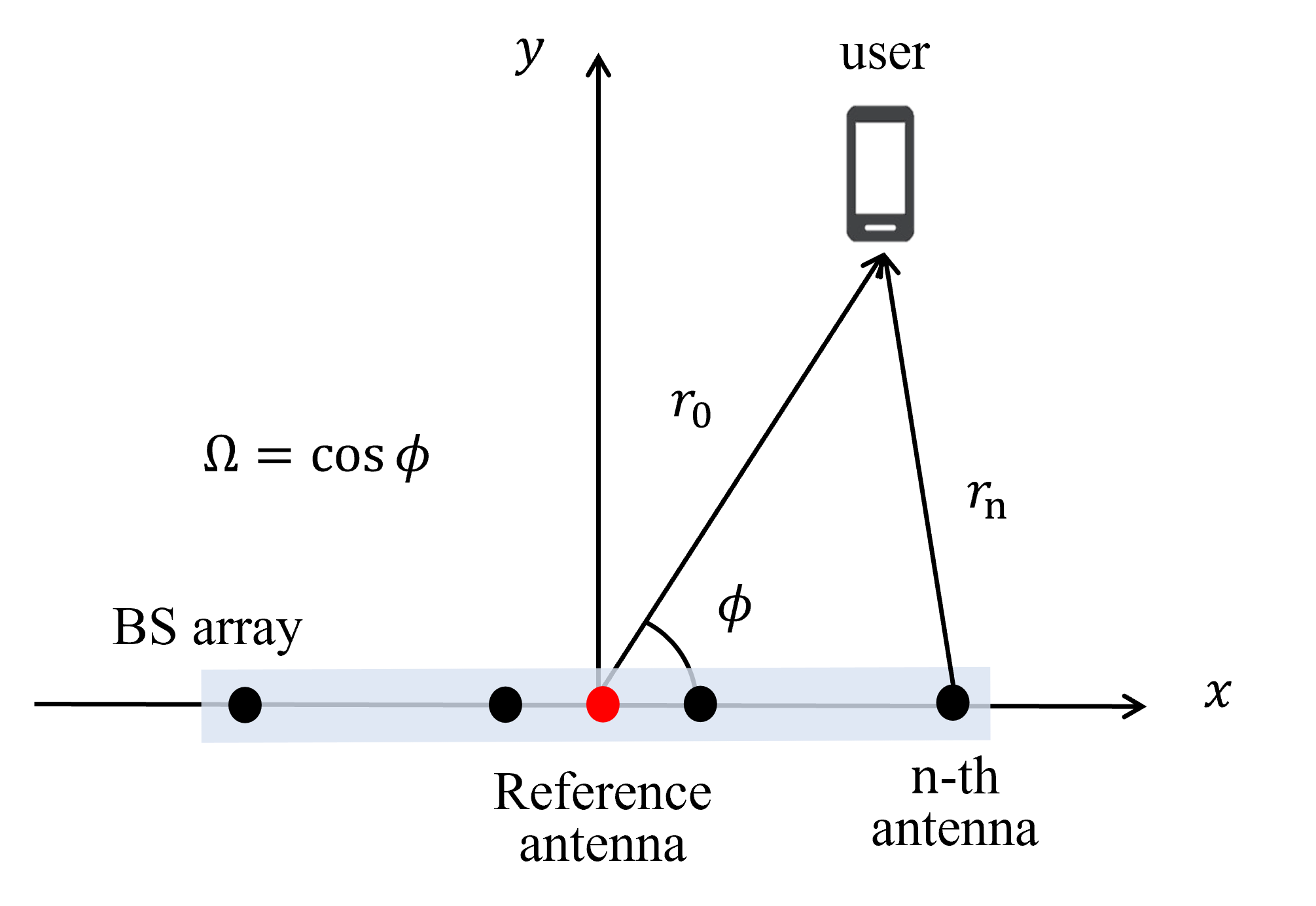}
\caption{Narrow-band XL-array system model}
\label{fig:nearfield-model}
\end{figure}

Fig. \ref{fig:nearfield-model} illustrates a narrow-band XL-array system, operating at a carrier frequency of $f_c$, with the carrier wavelength denoted as $\lambda=c/f_c$. The antenna spacing in the ULA is represented by $d$, In this paper, $d$ is set to $\lambda/2$, resulting in the array aperture of $D=(N-1)d=(N-1)\lambda/2$. The reference antenna is positioned at the center of the array, with the origin of the coordinate system chosen at the location of the reference antenna, and the x-axis aligned along the array direction. The distance and angle of departure (AOD) from the reference antenna to the user are denoted as $r_0$ and $\phi$, respectively.

The near-field range of the array is characterized by the Rayleigh distance \cite{raileigh-fresnel-dis}: $r_{ray}=2D^2/\lambda$. When $r_0<r_{ray}$, the user is within the near-field region of the BS, and the downlink channel can be modeled using the near-field steering vector \cite{linglongdai_spatial} as
\begin{equation}
\boldsymbol{h}_{near} = \sqrt{N}h_0\boldsymbol{b}(\Omega,r_0),
\label{eq:spctial_channel_vector}
\end{equation}
where $\boldsymbol{h}_{near}\in \mathbb{C}^{N\times1}$ is the channel vector and $h_0$ is the complex gain of the path at the reference antenna, $\Omega=\cos(\phi)$. Based on the spherical wavefront propagation model, considering both amplitude and phase variations, the near-field steering vector can be modeled as
\begin{equation}
\begin{aligned}
    \boldsymbol{b}(\Omega,r_0) = \frac{1}{\sqrt{N}}[\frac{r_0}{r_1}e^{-j\frac{2\pi}{\lambda}(r_1-r_0)},...,\frac{r_0}{r_N}e^{-j\frac{2\pi}{\lambda}(r_N-r_0)}]^T,
\end{aligned}
\label{eq:steering vector}
\end{equation}
where $r_n$ denotes the distance from the n-th antenna of BS array to the user. The x-coordinate of the n-th antenna is $x_n = d(n-\frac{N+1}{2})$, where $n\in \mathcal{N}= \left\{1,2,...,N\right\}$. Then $r_n$ can be calculated as $r_n=\sqrt{r_0^2+x_n^2-2r_0x_n\Omega}$.

By applying the DFT to the channel vector $\boldsymbol{h}_{near}$, the virtual angular domain representation \cite{virtual_channel} of the channel can be obtained as
\begin{equation}
\begin{aligned}
    \boldsymbol{H}_{near,A} &= \boldsymbol{F}^H\boldsymbol{h}_{near}\\
    &=\sum_{n=1}^{N}{\boldsymbol{a}^H(\Omega_n)\cdot \boldsymbol{h}_{near}}
\end{aligned}
\label{eq:angular domain vector}
\end{equation}
where $\boldsymbol{F} = [\boldsymbol{a}(\Omega_1),..,\boldsymbol{a}(\Omega_N)]\in \mathbb{C}^{N\times N}$ denotes the Fourier transform matrix. $\boldsymbol{a}(\Omega_n) = \frac{1}{\sqrt{N}}[1,e^{j\pi\Omega_n},...,e^{j(N-1)\pi\Omega_n}]^T$ denotes the far-field steering vector and $\Omega_n = \frac{2n-N-1}{N}$, $n\in \mathcal{N}$.

\subsection{Wave-Number Domain channel}

By extending $x_n$ to a continuous variable $x$, the near-field channel in Equation (1) can be transformed into a univariate function defined on $x\in \mathcal{X}=[-D/2,D/2]$ as follows:

\begin{equation}
\begin{aligned}
    h_{near}(x) = h_0\frac{r_0}{\sqrt{r_0^2+x^2-2r_0x\Omega}}e^{-j\frac{2\pi}{\lambda}(\sqrt{r_0^2+x^2-2r_0x\Omega}-r_0)}.
\end{aligned}
\label{eq:green function}
\end{equation}

By applying the FT to Equation (4), the near-field channel response can be expressed in the wave-number domain as $H_{near}(k_x)$, where the $k_x$ denotes the wave-number of x-axis:
\begin{equation}
\begin{aligned}
    H_{near}(k_x) = &\int_{-D/2}^{D/2} h_{near}(x)e^{-j k_x x}dx.
\end{aligned}
\label{eq:wave-number domain}
\end{equation}

The wave-number domain channel and angular domain channel are essentially the FT and DFT of the near-field channel response $h_{near}(x)$. Therefore, $\boldsymbol{H}_{near,A}$ can be viewed as a wave-number domain sampling of $H_{near}(k_x)$, while $\boldsymbol{h}_{near}$ can be viewed as a spatial domain sampling of $h_{near}(x)$. The sampling interval in the spatial domain is $d$, and only the portion of the wave-number domain with $|k_x|\leq \frac{2\pi}{\lambda}$ supports the transmission of communication power beyond the radiative near field \cite{HMIMO-wavenumber1}. Consequently, the wave-number domain bandwidth is denoted as $B_k = 2\cdot\frac{2\pi}{\lambda}$. Since $B_k\leq\frac{2\pi}{d}$, satisfying the Nyquist sampling theorem, $\boldsymbol{H}_{near,A}$ can be used to reconstruct $H_{near}(k_x)$.

Specifically, comparing the kernel function of (\ref{eq:wave-number domain}) with the exponential component of (\ref{eq:angular domain vector}), the sampling positions in the wave-number domain can be determined as $k_{x,n}=\frac{2\pi}{\lambda}\frac{2n-N-1}{N}$, and the sampling interval in the wave-number domain is denoted as $\Delta(k_x)=\frac{4\pi}{\lambda N}$. Consequently, $H_{near}(k_x)$ can be interpolated as
\begin{equation}
\begin{aligned}
    H_{near}(k_x) = d\sum_{n=-\infty}^{+\infty}{\boldsymbol{\tilde{H}}_{near,A}[n] \cdot \mathrm{sinc}\left(\frac{k_x-k_{x,n}}{\Delta(k_x)}\right)},
\end{aligned}
\label{eq:angle-to-wave}
\end{equation}
where $\boldsymbol{\tilde{H}}_{near,A}[n]$ denotes the $n$-th element of $\boldsymbol{\tilde{H}}_{near,A}$ and $\boldsymbol{\tilde{H}}_{near,A}$ is the periodic extension of $\boldsymbol{H}_{near,A}$, $\boldsymbol{\tilde{H}}_{near,A}[n]=\boldsymbol{H}_{near,A}[\text{mod}(n,N)]$. $\mathrm{sinc}(x) = \frac{\sin(\pi x)}{\pi x}$.

\section{Wave-Number Domain Approximate Expression}
\label{3}
In this section, the diffusion phenomenon of near-field channels in the wave-number domain and its approximate expression will be examined. Recent work \cite{youchangsheng_DFTcodebook} has indicated that the power diffusion range in the wave-number domain can be utilized for near-field beam training. Therefore, this section presents an approximate expression for the diffusion spectrum based on POSP. This approximation disregards the oscillation of the diffusion spectrum in the wave-number domain, while providing a closed-form relationship relating the diffusion range to both the angle and the distance of the user. Furthermore, by assuming that the user distance is not sufficiently close, a simplified expression for the diffusion range is derived.
\subsection{Closed-form Approximation Based on POSP}
By substituting the near-field single-end channel response (\ref{eq:green function}) into the Fourier transform in the wave-number domain (\ref{eq:wave-number domain}), the precise expression for the near-field angular domain channel can be obtained as
\begin{equation}
\setlength\abovedisplayskip{1pt}
\begin{aligned}
&H_{near}(k_x) = \int_{-\frac{D}{2}}^{\frac{D}{2}}dx \frac{r_0}{\sqrt{r_0^2+x^2-2r_0 x \Omega}}\cdot\\
&\exp\left\{j\left[-k_x x+\frac{2\pi}{\lambda}\left(r_0-\sqrt{r_0^2+x^2-2r_0 x \Omega}\right)\right]\right\}.\\
\end{aligned}
\label{eq:near_hkx}
\setlength\belowdisplayskip{0pt}
\end{equation}

The integrals shaped like (\ref{eq:near_hkx}) are known as oscillatory integral. It's difficult to obtain closed-form solution to (\ref{eq:near_hkx}) because of the presence of non-linear terms of $x$ in phase of the exponential part in (\ref{eq:near_hkx}), and the non-closed-form expression will make quantitative analysis challenging.
But it is worth noting that oscillatory integrals are prevalent in FM signal processing, and a method called POSP \cite{POSP-for-chirp} is widely used to approximate oscillatory integrals, which inspires us to take a similar approach to approximate (\ref{eq:near_hkx}).

\newtheorem{thm}{Theorem}

\begin{thm}[the Principle of Stationary Phase\cite{POSP-for-chirp}]
For an oscillatory integral of the form
\begin{equation}
I = \int_{\Omega}A(x)\exp\left\{j\psi(x)\right\}dx,
\label{eq:posp}
\end{equation}
where $A(x)>0$ and the first-order derivative function of $\psi(x)$ is continuous. 
When $A(x)$ changing slowly and $\psi(x)$ changing rapidly with $x$, the integrand tend to oscillate rapidly, the positive and negative values tend to cancel each other. Only the integrand near the stationary point $x_s$, i.e., the first-order derivative $\dot{\psi}(x_s)=0$, have contribution to $I$. Therefore, the approximation of $I$ can be written as 
\begin{equation}
I_a = \sqrt{\frac{2\pi}{|\ddot{\psi}(x_s)|}}A(x_s)\exp\left\{j\left[\psi(x_s)+\mathrm{sgn}[\ddot{\psi}(x_s)]\frac{\pi}{4}\right]\right\},
\label{eq:posp_app}
\setlength\belowdisplayskip{0pt}
\end{equation}
where $\ddot{\psi}(x)$ is the second-order derivative function of $\psi(x)$.
\end{thm}


By comparing (\ref{eq:near_hkx}) and (\ref{eq:posp}), the expressions of the magnitude and phase functions can be obtained as
\begin{equation}
\setlength\abovedisplayskip{0pt}
\begin{aligned}
&\quad\ 
A(x)=\frac{r_0}{\sqrt{r_0^2+x^2-2r_0x\Omega}},\\
\psi(x)=
&
-k_x x+\frac{2\pi}{\lambda}\left[r_0-\sqrt{r_0^2+x^2-2r_0x\Omega}\right],
\end{aligned}
\label{eq:A_phi_near_hkx}
\end{equation}
where $\psi(x)$ have unique stationary point
\begin{equation}
\setlength\abovedisplayskip{0pt}
\begin{aligned}
x_s=r_0\left[
\Omega-\frac{k_x}{\sqrt{\left(\frac{2\pi}{\lambda}\right)^2-k_x^2}}\sqrt{1-\Omega^2}
\right].
\end{aligned}
\label{eq:stationary_near_hkx}
\setlength\belowdisplayskip{0pt}
\end{equation}

Notice that the integral range of (\ref{eq:near_hkx}) is $\mathcal{X}=[-D/2,D/2]$, when $x_s\in \mathcal{X}$, $k_x$ needs to be satisfied by
\begin{equation}
\begin{aligned}
&k_x \in \mathcal{X}_k = \left[
\frac{\frac{2\pi}{\lambda}\left(\Omega-\frac{D}{2r_0}\right)}{\sqrt{1+(\frac{D}{2r_0})^2-\frac{D}{r_0}\Omega}},
\frac{\frac{2\pi}{\lambda}\left(\Omega+\frac{D}{2r_0}\right)}{\sqrt{1+(\frac{D}{2r_0})^2+\frac{D}{r_0}\Omega}}
\right].
\end{aligned}
\label{eq:kx_range}
\end{equation}

Therefore, when $k_x\in\mathcal{X}_k$, the approximation of (\ref{eq:near_hkx}) can be obtained according to (\ref{eq:posp_app}). When $k_x\notin\mathcal{X}_k$, the phase function $\psi(x)$ has no stationary points in the whole integral range $\mathcal{X}$, then the integral is approximated to be $0$. 

Thus, we can obtain the basic form of the approximate solution for (\ref{eq:near_hkx}): the diffusion spectrum defined over the finite interval $\mathcal{X}_k$ in the wave-number domain, with its amplitude and phase determined according to (\ref{eq:posp_app}), (\ref{eq:A_phi_near_hkx}), (\ref{eq:stationary_near_hkx}) and (\ref{eq:kx_range}). By examining the expression for $\mathcal{X}_k$, it can be observed that the power diffusion range is distributed around $k_x = \frac{2\pi}{\lambda}\Omega$ and it is always wider on the side where $|k_x|$ is smaller. As the user distance $r_0$ decreases, the diffusion width increases, and this width is also related to the user's azimuth angle $\Omega$.

\subsection{Simplified Expression}

Although (\ref{eq:kx_range}) provides a closed-form expression for the diffusion range $\mathcal{X}_k = [k_l, k_r]$ when given the user position $(r_0, \Omega)$. However, in practical applications, it is often necessary to estimate $(r_0, \Omega)$ based on $\mathcal{\tilde{X}}_k = [k_l, k_r]$, where $\mathcal{\tilde{X}}_k$ can be defined as $\mathcal{\tilde{X}}_k = \left\{k_x : |H_{near}(k_x)|<\beta \max|H_{near}(k_x)|\right\}$ ($\beta$ is a given constant, whose specific value will influence the performance of the approximation model. This will be further discussed in Section. \ref{4}). This objective is challenging to achieve through (\ref{eq:kx_range}). 

However, it is noteworthy that $D/r_0$ appears multiple times in (\ref{eq:kx_range}).
In practical scenarios, the distance between the users and BS is generally not small enough to be on the order of the array aperture; hence, $r_0\gg D$, by taking the Taylor series expansion as $D/r_0\ll1$, and retaining the first-order term, $\mathcal{X}_k$ can be approximate as

\begin{equation}
\begin{aligned}
&\mathcal{X}_{k}\approx\mathcal{X}_{k,s}=\\
&\left[\frac{2\pi}{\lambda}\left(\Omega-\frac{D}{2r_0}(1-\Omega^2)\right),
\frac{2\pi}{\lambda}\left(\Omega+\frac{D}{2r_0}(1-\Omega^2)\right)
\right].
\label{eq:simp_omega}
\end{aligned}
\end{equation}

The $\mathcal{X}_{k,s}$ is symmetric about $\frac{2\pi}{\lambda}\Omega$, with a width of $\frac{2\pi D (1-\Omega^2)}{\lambda r_0}$. Therefor, The user's $\Omega$ and $r_0$ can be estimated by $k_l,k_r$ as
\begin{equation}
\begin{aligned}
\widetilde{\Omega} = \frac{\lambda}{4\pi}(k_l+k_r), \widetilde{r}_0 = \frac{2\pi D}{\lambda} \frac{1-\widetilde{\Omega}^2}{k_r-k_l}.
\label{eq:user_r_omega}
\end{aligned}
\end{equation}

It is worth noting that the approximations given in equations (\ref{eq:simp_omega}) and (\ref{eq:user_r_omega}) are not applicable in the far-field scenario. Specifically, as $r_0\rightarrow r_{ray}$, the amplitude of $h_{near}(x)$ approximates to 1, and the phase of $h_{near}(x)$ can be approximated using a Taylor expansion, retaining only the first-order term with respect to $x$. In this case, (\ref{eq:near_hkx}) asymptotically approaches the far-field wave-number domain channel $H_{far}(k_x)$:
\begin{equation}
\begin{aligned}
H_{far}(k_x) &= \int_{-\frac{D}{2}}^{\frac{D}{2}} \exp\left\{j\left(-k_x x+\frac{2\pi}{\lambda}x \Omega \right)\right\}dx\\
& = \frac{2\sin[(\frac{D}{2}(k_x-\frac{2\pi}{\lambda}\Omega)]}{k_x-\frac{2\pi}{\lambda}\Omega}.
\end{aligned}
\label{eq:far_hkx}
\end{equation}
According to (\ref{eq:far_hkx}), as $r_0$ increases, the diffusion width does not continuously decrease but asymptotically approaches a certain value. Therefore, when the width given by (\ref{eq:simp_omega}) is smaller than the 3dB width $B_{3dB}=0.866\cdot\frac{2\pi}{D}$ of the main lobe of $H_{far}(k_x)$, it is considered that the approximation model will fail, namely:

\begin{equation}
\begin{aligned}
0.866\cdot\frac{2\pi}{D}&>\frac{2\pi D(1-\Omega^2)}{\lambda r_0}\\
r_0 &>1.155\cdot\frac{D^2(1-\Omega^2)}{\lambda}.
\end{aligned}
\label{eq:eff_ray}
\end{equation}

Comparing the right side of (\ref{eq:eff_ray}) and $r_{ray}$, the right side of (\ref{eq:eff_ray}) includes an additional $1-\Omega^2=\sin^2(\phi)$ factor, indicating that the near-field range is not equidistant in all directions. This phenomenon has also been mentioned in \cite{eff_ray}, where the near-field boundary shaped like (\ref{eq:eff_ray}) is also called effective Rayleigh distance.

\section{Simulations and Results}
\label{4}
\vspace{-1mm}

In this section, the accuracy of the proposed model are evaluated through simulation, and a beam training scheme is presented as an example to illustrate the practical applications of the proposed model.
First, numerical simulations of the wave-number domain channel is presented to visually illustrate the diffusion phenomenon of the near-field channel and the intuitive approximation effects of the proposed model on this diffusion.
Secondly, the accuracy of the proposed model's approximation of the diffusion range is compared, as this directly impacts the model's potential applications.
Finally, based on the beam training scheme proposed in \cite{youchangsheng_DFTcodebook}, an improved scheme is introduced using the wave-number domain model as an example of the potential application of the proposed model. The effectiveness of the algorithm is evaluated using the NMSE of angle and distance estimation and achievable rate of beam training.
In the simulation, without further elaboration, the ULA has $N=256$ antennas and the spacing between antennas is $\lambda/2$. The carrier frequency $f_c=30$ GHz, so the Rayleigh distance of the ULA is $r_{ray}=327.68$ m.

\subsection{Wave-Number Domain Numerical Results}

\begin{figure}[htbp]
\centering
    \centering
    \includegraphics[width=0.4\textwidth]{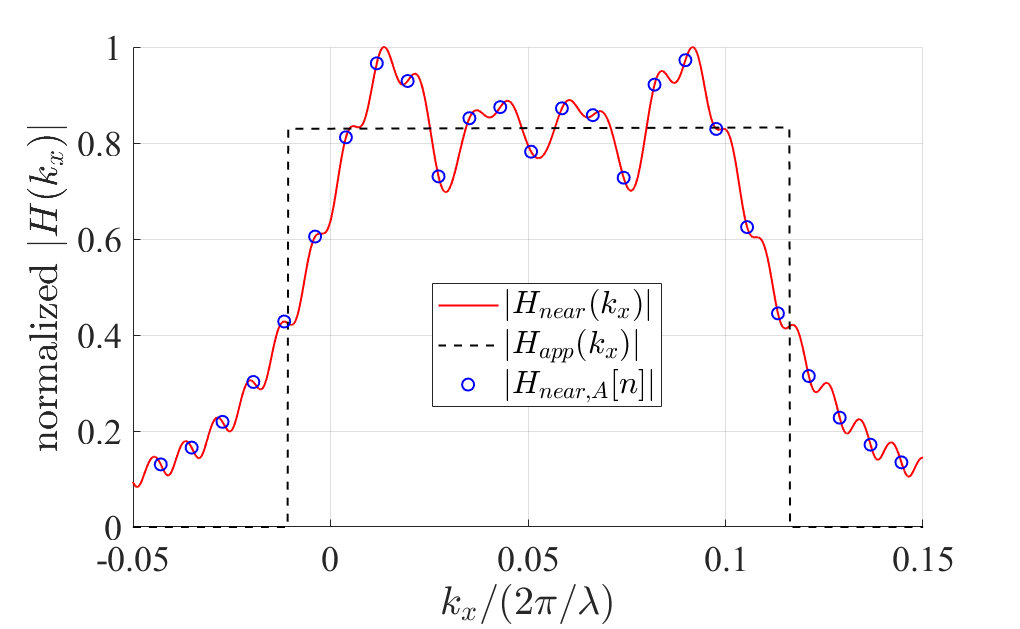}
    \caption{The simulation of the normalized amplitude of wave-number domain channel, wave-number domain approximation and angular domain channel.}
    \label{fig:approx_P}
\end{figure}

Firstly, we simulate the near-field wave-number domain channel $|H_{near}(k_x)|$ given by (\ref{eq:near_hkx}), the proposed approximation $|H_{app}(k_x)|$ given by (\ref{eq:posp_app}), (\ref{eq:A_phi_near_hkx}), (\ref{eq:stationary_near_hkx}) and (\ref{eq:kx_range}).
The angular domain channel $|\boldsymbol{H}_{near,A}|$ given by (\ref{eq:angular domain vector}) is also simulated, shown as Fig. \ref{fig:approx_P}.

Fig. \ref{fig:approx_P} shows the wave number domain amplitude spectrum when the user position is $r_0=10$ m, $\Omega=0.05$, normalized using the maximum value of $|H_{near}(k_x)|$. As can be seen from the figure, the wave-number domain power appears to diffuse with $k_x = \frac{2\pi}{\lambda}\Omega$ as the center, $|H_{app}(k_x)|$ can approximate the basic form of $|H_{near}(k_x)|$ while neglecting its oscillatory effects. The angular domain spectrum is formally a sample of the wave-number domain spectrum. However, since the angular domain is defined on a set of grids, the width of the diffusion spectrum cannot be accurately estimated.

\subsection{Accuracy of Diffusion Range}

Next, we compare the accuracy of the proposed approximate model in estimating the diffusion width through simulations. Two approximations for the diffusion range $\mathcal{X}_k$ are presented in (\ref{eq:kx_range}) and $\mathcal{X}_{k,s}$ are presented in (\ref{eq:simp_omega}). In the wave-number domain approximation, the diffusion range is defined as $\mathcal{\tilde{X}}_k=\left\{k_x:|H_{near}(k_x)|<\beta \max|H_{near}(k_x)|\right\}$. From Fig. \ref{fig:approx_P}, it can be observed that the intersection of $H_{app}(k_x)$ and $H_{near}(k_x)$ is roughly located at half of the maximum value of $H_{app}(k_x)$. However, in practical applications, only $H_{near}(k_x)$ is obtainable, so the maximum value of $H_{near}(k_x)$ must be used to substitute for the maximum value of $H_{app}(k_x)$. Since $H_{app}(k_x)$ neglects the fluctuations of $H_{near}(k_x)$, the parameter $\beta$ should be chosen to be slightly less than 0.5. In simulation, $\beta$ is set to 0.42.

\begin{figure}[htbp]
    \centering
    \subfigure[]{\includegraphics[width=0.4\textwidth]{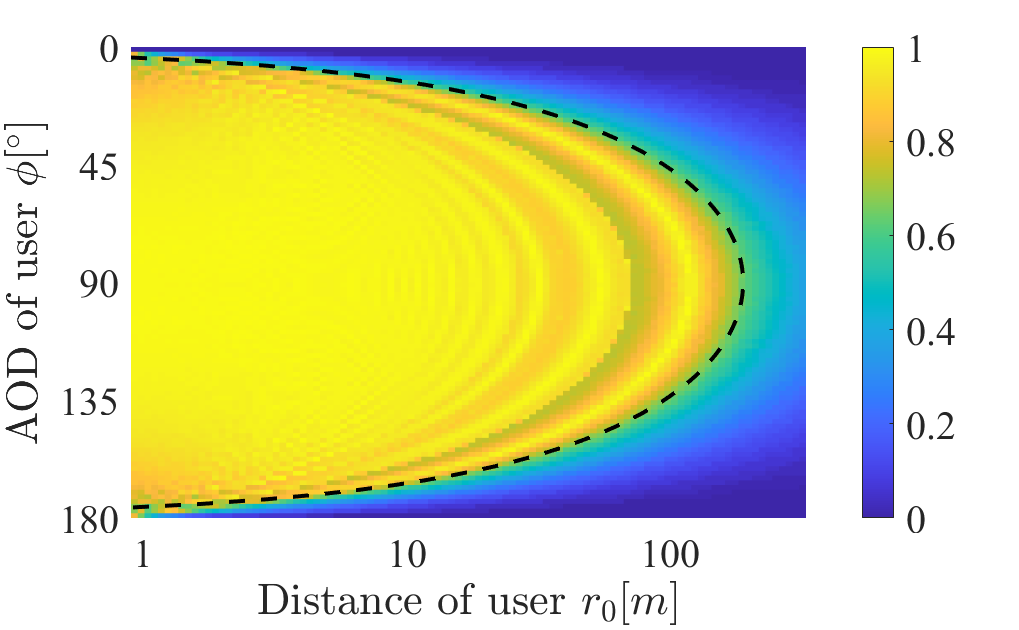}\label{fig:jaccard_figure1}}
    \subfigure[]{\includegraphics[width=0.4\textwidth]{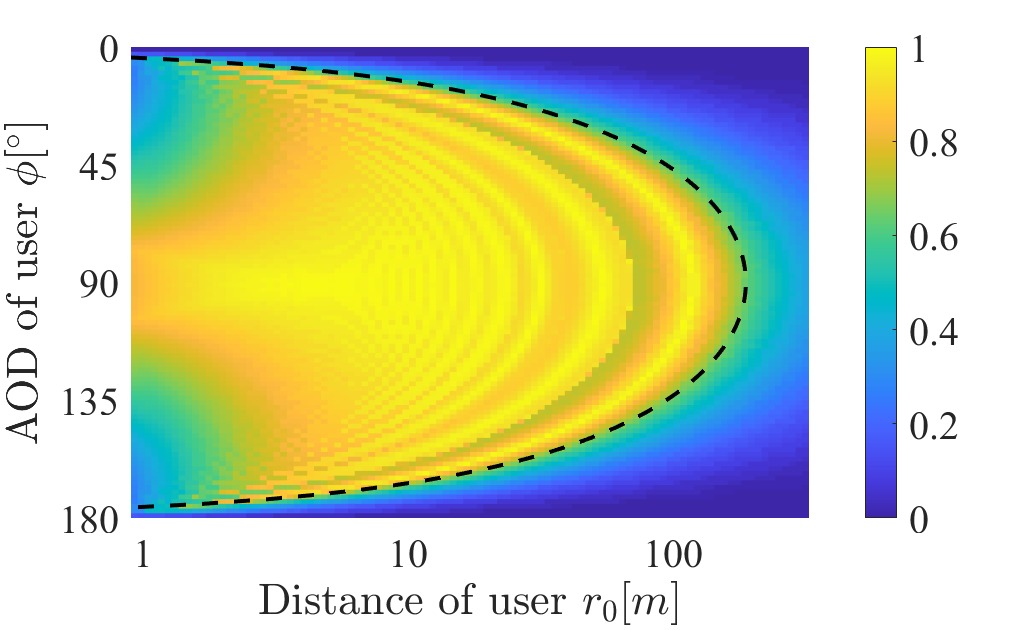}\label{fig:jaccard_figure2}}
    \caption{Simulation of Jaccard Index: (a) between $\mathcal{X}_k$ and baseline. (b) between $\mathcal{X}_{k,s}$ and baseline. The black dashed line marks the near-field region determined by the effective Rayleigh distance, which is given by (\ref{eq:eff_ray})}
    \label{fig:jaccard}
\end{figure}

To measure the similarity between two intervals $A$ and $B$, we use the Jaccard Index, which can be defined as
\begin{equation}
\begin{aligned}
J_{AB} = \frac{|A\bigcap B|}{|A\bigcup B|},
\end{aligned}
\label{eq:jaccard}
\end{equation}
where $|A|$ denotes the length of the interval $A$. The range of $J_{AB}$ is $[0,1]$, and The larger $J_{AB}$ is, the more similar the two intervals $A$ and $B$ are. We used $\mathcal{\tilde{X}}_k$ as baseline and simulate the Jaccard index between $\mathcal{X}_k$ and the baseline, $\mathcal{X}_{k,s}$ and the baseline, respectively, as shown in Fig. \ref{fig:jaccard}.

As shown in Fig. \ref{fig:jaccard_figure1}, $\mathcal{X}_k$ exhibits high accuracy within the effective Rayleigh distance, with a Jaccard index larger than $0.73$. The smaller the user distance $r_0$ is, the higher the Jaccard index. In Fig. \ref{fig:jaccard_figure2}, when the user distance $r_0$ is larger, the jaccard index of $\mathcal{X}_{k,s}$ is similar to that of $\mathcal{X}_k$. However, when the user distance $r_0$ approaches the order of $D$, the accuracy of the approximation significantly decreases, as the condition $D/r_0\leq 1$ is no longer satisfied, which is consistent with the previous theoretical analysis in section \ref{3}. In practical applications, if it can be assumed that the user distance is always larger than the order of $D$, then $\mathcal{X}_{k,s}$ can be used to approximate $\mathcal{\tilde{X}}_k$.

\subsection{Potential Applications of the Proposed Approximation: A Beam Training Example}

We discuss a potential application of the proposed approximation model using the fast near-field beam training scheme presented in reference \cite{youchangsheng_DFTcodebook}: the angle support width based joint angle and range estimation (ASW-JE) scheme. The main idea of the ASW-JE scheme is to utilize a far-field codebook for beam sweeping to obtain the angular domain channel as shown in equation (\ref{eq:angular domain vector}). Subsequently, based on the characteristics of the diffusion width of the angular domain channel, the user's AOD and distance are estimated.

The proposed model can further enhance the performance of the ASW-JE beam training scheme. We name the improved scheme as the wave-number domain support width based joint angle and range estimation (WDSW-JE) scheme. Specifically, the angular domain channel $\boldsymbol{H}_{near,A}$ obtained through far-field beam sweeping is used to reconstruct the wave-number domain channel $H_{near}(k_x)$ based on equation (\ref{eq:angle-to-wave}). Subsequently, the diffusion interval $\mathcal{\tilde{X}}_k$ is estimated in the wave-number domain, and the user's AOD $\widetilde{\Omega}$ and distance $\widetilde{r}_0$ are estimated based on (\ref{eq:user_r_omega}). Finally, the optimal codebook $\boldsymbol{b}(\widetilde{\Omega},\widetilde{r}_0)$ is generated based on the estimated AOD and distance.

We validate the effectiveness of the proposed scheme through simulations. The following beam training schemes are considered in the simulations: (1) Exhaustive-search beam training \cite{linglongdai_spatial}: This method sweeps the entire two-dimensional polar-domain codebook to obtain the optimal codebook. (2) ASW-JE scheme: As shown in \cite{youchangsheng_DFTcodebook}, We consider the cases of K=3, which further examines three adjacent candidate angles for secondary beam training after estimating the user's AOD and distance. (3) WDSW-JE: This scheme is an improved ASW-JE based on our proposed wave-number domain approximation model. (4) Perfect CSI: This approach directly generates the optimal codebook $\boldsymbol{b}(\Omega,r_0)$ based on the prior user's location.

In the simulations, we conduct $1000$ random experiments under different signal-to-noise ratio (SNR) conditions. The user distance is set to $r_0 = 20m$, while the user angle $\Omega$ follows a uniform distribution over the range $[-1,1]$. We first compare the performance of different schemes in estimating the user's position $(\Omega, r_0)$. The normalized mean square error (NMSE) is employed to evaluate the accuracy of estimated angle and distance,which can be defined as NMSE$_{angle} = \frac{\mathbb{E}\left(|\widetilde{\Omega}-\Omega|^2\right)}{\mathbb{E}\left(|\Omega|^2\right)}$ and NMSE$_{distance} = \frac{\mathbb{E}\left(|\widetilde{r}_0-r_0|^2\right)}{\mathbb{E}\left(|r_0|^2\right)}$.
Then we use achievable rate to compare the performance of different schemes. The achievable rate is calculated as $R=\log_{2}\left(1+\frac{|\boldsymbol{h}_{near}^H \boldsymbol{v}|}{\sigma ^2}\right)$, where $\boldsymbol{h}_{near}$ represents the user’s channel vector and $\boldsymbol{v}$ is the beamforming vector provided by each scheme. To account for the impact of overhead in beam training, we use the effective achievable rate defined in \cite{youchangsheng_DFTcodebook} as $R_{eff}=\left(1-\frac{T_{tra}}{T_{tot}}\right)R$, which incorporates the training overhead. $T_{tot}$ denotes the total number of symbols in each time frame and $T_{tra}$ denotes the number of required training symbols.

\begin{figure}[htbp]
    \centering
    \subfigure[]{
    \includegraphics[width=0.4\textwidth]{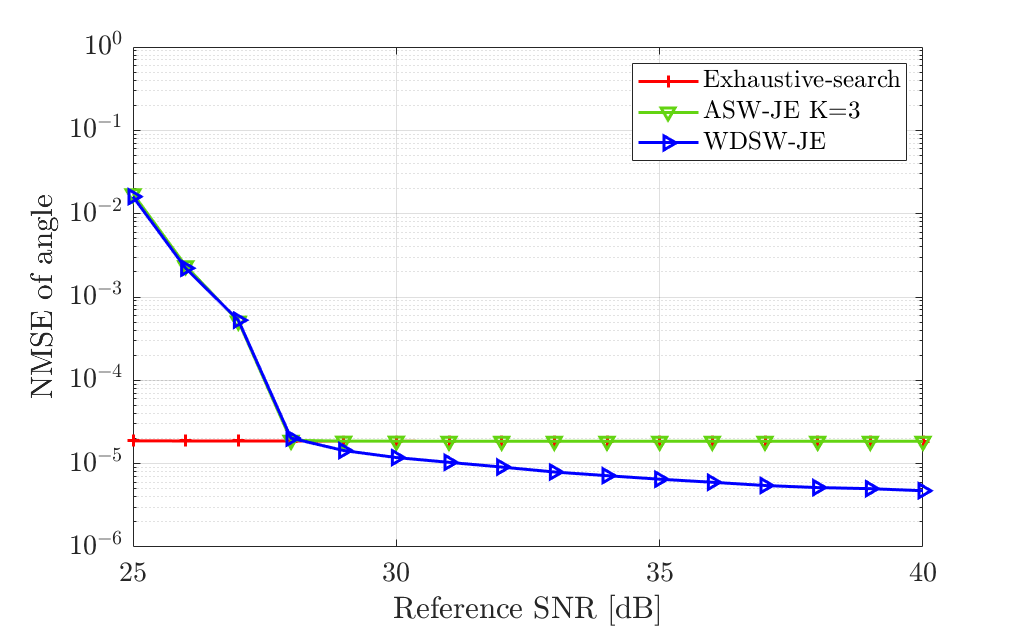}\label{fig:NMSE_angle}
    }
    \subfigure[]{\includegraphics[width=0.4\textwidth]{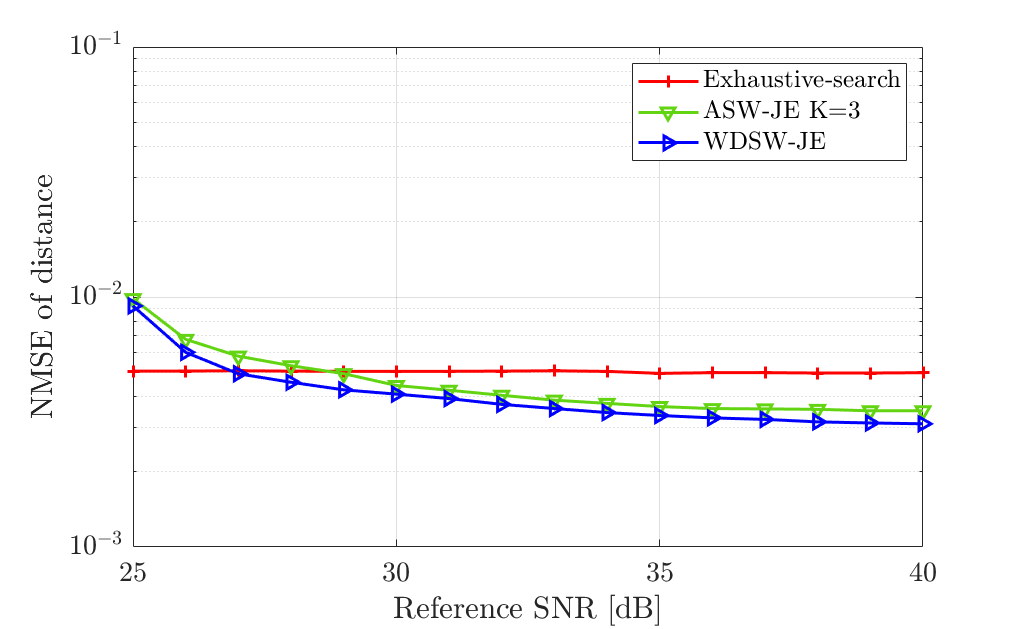}\label{fig:NMSE_r}}
    \caption{Estimation NMSE versus reference SNR
     (a) NMSE of user AOD, $\Omega$, (b) NMSE of user distance, $r_0$}
    \label{fig:NMSE}
\end{figure}

In Fig. \ref{fig:NMSE}, the estimated NMSE of the WDSW-JE scheme is lower than the existing ASW-JE schemes. At higher SNR, it even outperforms the exhaustive search scheme. The WDSW-JE method achieves a lower NMSE than the exhaustive search because the exhaustive codebook is defined on a series of angle-distance grids, whereas the wave-number domain approach surpasses the angular grid limitation, obtaining higher resolution through oversampling.

\begin{figure}[htbp]
    \centering
    \subfigure[]{
    \includegraphics[width=0.4\textwidth]{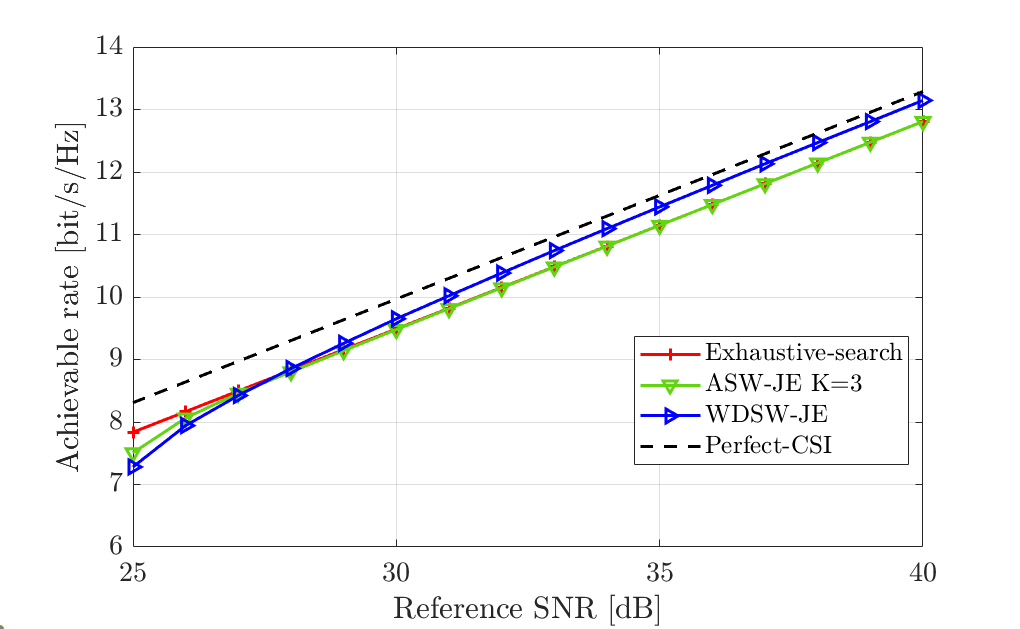}\label{fig:achievable_rate}
    }
    \subfigure[]{\includegraphics[width=0.4\textwidth]{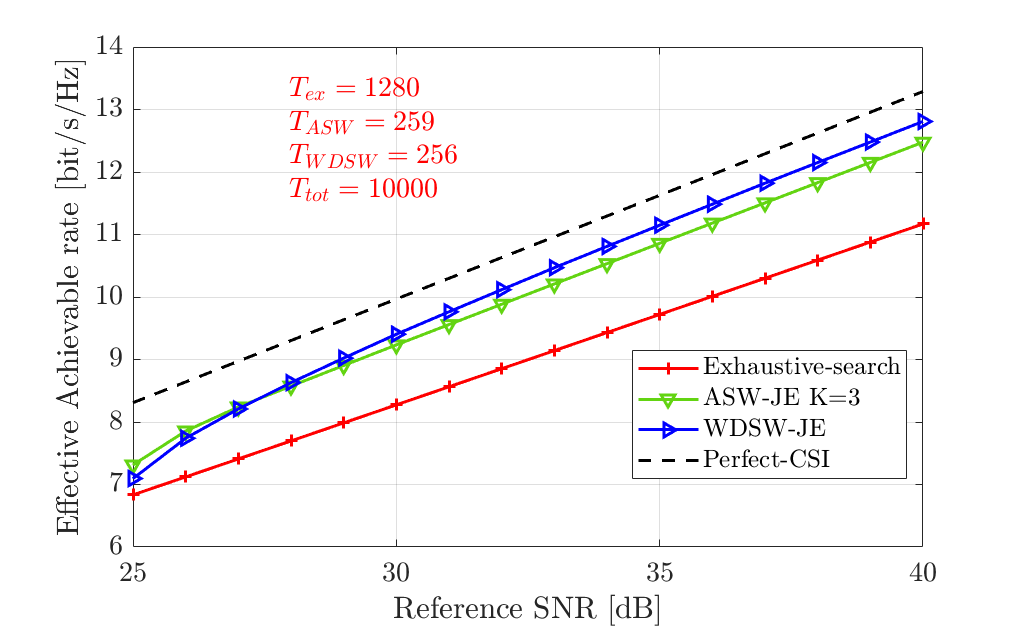}\label{fig:eff_achievable_rate}}
    \caption{Achievable rate versus reference SNR
     (a) Achievable rate, (b) Effective achievable rate.}
    \label{fig:rate}
\end{figure}

Fig. \ref{fig:rate} presents the simulation results of the achievable rate for the user. In Fig. \ref{fig:achievable_rate}, under high SNR conditions, both the ASW-JE and WDSW-JE schemes achieve a higher achievable rate than the exhaustive search method, with WDSW-JE exhibiting an even higher rate. This is because both schemes enable off-grid estimation in the distance dimension, while WDSW-JE also allows off-grid estimation in the wave-number domain. At low SNR, the achievable rate is slightly lower compared to the exhaustive search scheme due to drastical fluctuations in the received wave-number domain spectrum, which introduces errors in angle estimation \cite{youchangsheng_DFTcodebook}.
In Fig. \ref{fig:eff_achievable_rate}, the impact of overhead on various methods is further considered. It can be observed that WDSW-JE, with its lower overhead and higher estimation accuracy, achieves a higher effective achievable rate.

\section{Conclusion}
\label{5}
In this paper, we discusse the expression of near-field wave-number domain channel and its approximation methods. 
We first presente a closed-form approximation based on POSP, which ignores the oscillatory effects of the wave-number domain channel and reflects its diffusion range. Then we propose a simplified approximation under the condition that the user distance is larger than the array aperture.
Subsequently, we validate the accuracy of the diffusion range approximation through simulations: within the effective Rayleigh distance, the closed-form model demonstrates high accuracy. Meanwhile, the simplified approximation also exhibits favorable approximation performance when the user distance exceeds the order of magnitude of the array aperture.
Finally, we explored a potential application of the model: a beam training scheme. Based on the wave-number domain approximation, we improved the existing ASW-JE scheme, achieving higher user location estimation accuracy and higher achievable rates.

\bibliographystyle{IEEEtran}
\bibliography{bib}
\end{document}